\documentclass[10pt,a4paper]{article}
\usepackage{amsmath}
\usepackage{exscale}
\usepackage{graphics}
\usepackage[dvips]{graphicx}
\usepackage{latexsym}
\usepackage{amssymb}
\usepackage{overcite}
\usepackage{klucite}
\usepackage[latin1]{inputenc}
\usepackage[section]{placeins}
\usepackage[top=2.5cm,bottom=2.5cm,left=2.5cm,right=2.1cm]{geometry}

\title{Comments on: "Impossibility of the existence of the universal density functional" V.B.Vobrov, S.A. Trigger. arXiv: 1012.3241 v1. Dec 2010}

\author{Ramón Alain Miranda Quintana $^{\S}$, Daniel Codorniu Pujals}

\begin{document}

\maketitle

\vspace{1cm}
\begin{small}
$\S$ email: alain@instec.cu  "Higher Institute of Technology and Applied Sciences, Havana, Cuba"
\end{small}

\vspace{1cm}

The paper is based in the original formulation of DFT presented by Hohenberg and Kohn\cite{Hohenberg}. In that paper the authors dealt only with systems which ground states are non-degenerated, using this condition in order to obtain most of the outcomes presented in the cited work. As a consequence, some results are either not presented or derived in a general form. For example, they defined the functional $F_{HK}[\rho]$  as $F_{HK}[\rho]=\langle\psi_{o}[\rho]|T+V|\psi_{o}[\rho]\rangle$, as it is also done in the first part of the expression (4), been this an expression that is only valid for non-degenerated ground states.  The general form of the $F_{HK}[\rho]$ functional is the one presented in the second part of expression (4):
$$F_{HK}[\rho]=E_{o}(\{\rho_{o}\})-\int\varphi^{ext}(r)\rho(r)dr$$
been this the form that can be used for any ground state.

It is true that Hohemberg and Kohn assumed, without any demonstration, that the above mentioned functional is universal, in the sense that it is independent of the external potential \cite{Hohenberg}. It is also certain that in many derivations of the variational principle based on the electronic density, that can be found in the literature, the universality of the $F_{HK}[\rho]$ is used without no previous demonstration \cite{Koch,Parr}. But, the essential point here, is that the universality of the functional   $F_{HK}[\rho]$ is not necessary in order to construct such variational principle. In the book "The Fundamentals of Density Functional Theory" \cite{Eschrig}, the author demonstrated that, no matter the functional $F_{HK}[\rho]$ is universal or not, a variational principle based in the electronic density can be rigorously stated.

Nevertheless, we have the fundamental question about the universality of the $F_{HK}[\rho]$ . This is a very important question, not only form the theoretical point of view, but also for practical reasons, because, in the hypothetical case that $F_{HK}[\rho]$ was not universal, then we are obliged to find a different functional for each concrete problem (i.e. external potential).  Fortunately, the answer to this crucial question was given by Levy and Lieb \cite{Levy,Lieb}. In order to solve the v-representability  problem of $F_{HK}[\rho]$, these authors defined a new functional:$F_{LL}[\rho]=inf_{\psi\rightarrow\rho}\langle\psi|T+V|\psi\rangle$, here the infimum search is over all N-particle wave functions (not only ground states) yielding a given density $\rho$. In this way, besides having a functional whose domain of definition is the set of all the N-representable densities, it is obtained a universal functional because in its definition it wasn't made any reference to the external potential. So, the question of the universality of the functional $F_{HK}[\rho]$ is answered through its extension, $F_{LL}[\rho]$ , over all N-representable densities.

\clearpage
\addcontentsline{toc}{section}{References} %ok

\end{document}